\documentclass[sigconf]{acmart}

\usepackage{booktabs} 
\usepackage{todonotes}
\usepackage{microtype}
\usepackage{xspace}
\usepackage{balance}
\usepackage{algorithm}
\usepackage{algpseudocode}
\usepackage{caption}
\usepackage[compact]{titlesec}
\usepackage{multicol} 
\usepackage{subfig}

\usepackage{pifont}

\usepackage{tikz}
\usepackage{pgflibraryshapes}
\usetikzlibrary{arrows}
\usetikzlibrary{chains}
\usetikzlibrary{fadings}
\usetikzlibrary{matrix}
\usetikzlibrary{fit}
\usetikzlibrary{positioning}
\usetikzlibrary{shapes}
\usetikzlibrary{automata}
\tikzset{
 ctrlstate/.style = {state,align=center,inner sep=2pt, minimum size=2mm},
 cfastate/.style = {ctrlstate},
 cfatargetstate/.style = {cfastate, double},
 compstate/.style = {cfastate, rounded rectangle, minimum height=5mm, minimum width=3em, inner sep=3pt},
 concept/.style = {cfastate, inner sep=3pt, fill=gray!50, rectangle,
        minimum width=17mm, minimum height=9mm, draw=gray!6 },
 crosscutting/.style = {concept, rounded rectangle, fill=gray!30},
 conceptstate/.style = {concept, rounded rectangle, fill=gray!30, draw=gray!90, minimum width=20mm},
 inputstate/.style = {concept, rectangle, fill=gray!10, draw=gray!90, minimum width=20mm, inner sep=3pt},
 explain/.style = {circle, draw=gray!30, line width=1mm, minimum size=7mm},
 one/.style = {fill=blue!70,draw=blue!70},
 two/.style = {fill=red!50,draw=red!50},
 abststate/.style = {rectangle,align=center,inner sep=2pt,minimum size=3.5mm,fill=gray!0, draw=gray!90},
 line/.style = {draw},
 trans/.style = {draw,semithick,->,shorten >=1pt,>=stealth'},
 missing/.style = {draw=red,densely dotted,fill=red,semithick,->,shorten >=1pt,>=stealth'},
 ctrans/.style = {draw,very thick,->,shorten >=1pt,>=stealth',draw=gray!90},
 epsilon/.style = {trans,dashed},
 strengthen/.style = {draw=gray!30,semithick,double,shorten >=1pt,>=stealth',line width=1mm},
}

\newcommand{\motionblockleft}{\begin{mbox}\sf\begin{tikz}[baseline=(X.base)]\node[draw=black!60,fill=blue!12,semithick,rectangle,inner sep=1pt, minimum size=1em, outer sep=0pt, rounded corners=1pt] (X)}%
\newcommand{\controlblockleft}{\begin{mbox}\sf\begin{tikz}[baseline=(X.base)]\node[draw=black!60,fill=orange!15,semithick,rectangle,inner sep=1pt, minimum size=1em, outer sep=0pt, rounded corners=1pt] (X)}%
\newcommand{\hatblockleft}{\begin{mbox}\sf\begin{tikz}[baseline=(X.base)]\node[draw=black!60,fill=yellow!20,semithick,rectangle,inner sep=1pt, minimum size=1em, outer sep=0pt, rounded corners=1pt] (X)}%
\newcommand{\looksblockleft}{\begin{mbox}\sf\begin{tikz}[baseline=(X.base)]\node[draw=black!60,fill=violet!20,semithick,rectangle,inner sep=1pt, minimum size=1em, outer sep=0pt, rounded corners=1pt] (X)}%
\newcommand{\sensingblockleft}{\begin{mbox}\sf\begin{tikz}[baseline=(X.base)]\node[draw=black!60,fill=cyan!20,semithick,rectangle,inner sep=1pt, minimum size=1em, outer sep=0pt, rounded corners=1pt] (X)}%
\newcommand{\soundblockleft}{\begin{mbox}\sf\begin{tikz}[baseline=(X.base)]\node[draw=black!60,fill=magenta!20,semithick,rectangle,inner sep=1pt, minimum size=1em, outer sep=0pt, rounded corners=1pt] (X)}%
\newcommand{\operatorblockleft}{\begin{mbox}\sf\begin{tikz}[baseline=(X.base)]\node[draw=black!60,fill=green!20,semithick,rectangle,inner sep=1pt, minimum size=1em, outer sep=0pt, rounded corners=1pt] (X)}%

\newcommand{\blockleft}{\begin{mbox}\sf\begin{tikz}[baseline=(X.base)]\node[draw=black!60,fill=black!3,semithick,rectangle,inner sep=1pt, minimum size=1em, outer sep=0pt, rounded corners=1pt] (X)}%
\newcommand{\blockright}{;\end{tikz}\normalfont\end{mbox}}%

\newcommand{\motionblock}[1]{\motionblockleft{\small\textsf{#1}}\blockright}
\newcommand{\controlblock}[1]{\controlblockleft{\small\textsf{#1}}\blockright}
\newcommand{\hatblock}[1]{\hatblockleft{\small\textsf{#1}}\blockright}
\newcommand{\looksblock}[1]{\looksblockleft{\small\textsf{#1}}\blockright}
\newcommand{\sensingblock}[1]{\sensingblockleft{\small\textsf{#1}}\blockright}
\newcommand{\soundblock}[1]{\soundblockleft{\small\textsf{#1}}\blockright}
\newcommand{\operatorblock}[1]{\operatorblockleft{\small\textsf{#1}}\blockright}

\newcommand{\mblockleft}{\begin{mbox}\sf\begin{tikz}[baseline=(X.base)]\node[draw=red!60,densely dotted,fill=red!3,semithick,rectangle,inner sep=1pt, minimum size=1em, outer sep=0pt, rounded corners=1pt] (X)}%
\newcommand{\mblockright}{;\end{tikz}\normalfont\end{mbox}}%

\renewcommand{\paragraph}[1]{\vspace{0.08em}\noindent {\bf #1}}

\newcommand{\solutionpattern}{{\Large$\spadesuit$}\xspace}
\newcommand{\elementarypattern}{{\Large$\heartsuit$}\xspace}
\newcommand{\evidencevariables}{{\Large$\clubsuit$}\xspace}
\newcommand{\gameprogramming}{{\Large$\diamondsuit$}\xspace}

\AtBeginDocument{%
  \providecommand\BibTeX{{%
    \normalfont B\kern-0.5em{\scshape i\kern-0.25em b}\kern-0.8em\TeX}}}

\copyrightyear{2021}
\acmYear{2021}
\setcopyright{acmcopyright}
\acmConference[WiPSCE '21]{The 16th Workshop in Primary and Secondary Computing Education}{October 18--20, 2021}{Virtual Event, Germany}
\acmBooktitle{The 16th Workshop in Primary and Secondary Computing Education (WiPSCE '21), October 18--20, 2021, Virtual Event, Germany}
\acmPrice{15.00}
\acmDOI{10.1145/3481312.3481346}
\acmISBN{978-1-4503-8571-8/21/10}

\newtheorem{definition}{Definition}

\newcommand{\summary}[2]{
	\vspace{0.4em}
	\noindent
	\colorbox{gray!20}{%
		\parbox{.97\linewidth}{%
			\textbf{\textsf{Summary (\textit{#1})}}
			#2
		}%
	}%
}%

\begin{document}

\title{Code Perfumes: Reporting Good Code to Encourage Learners}

\author{\mbox{Florian Obermüller}}
\email{obermuel@fim.uni-passau.de}
\affiliation{%
	\institution{University of Passau}
	\city{Passau}
	\country{Germany}
}

\author{Lena Bloch}
\email{blochl@fim.uni-passau.de}
\affiliation{%
	\institution{University of Passau}
	\city{Passau}%
	\country{Germany}
}

\author{Luisa Greifenstein}
\email{luisa.greifenstein@uni-passau.de}
\affiliation{%
	\institution{University of Passau}
	\city{Passau}%
	\country{Germany}
}

\author{Ute Heuer}
\email{ute.heuer@uni-passau.de}
\affiliation{%
	\institution{University of Passau}
	\city{Passau}%
	\country{Germany}
}

\author{Gordon Fraser}
\email{gordon.fraser@uni-passau.de}
\affiliation{%
	\institution{University of Passau}
	\city{Passau}
	\country{Germany}
}
\renewcommand{\shortauthors}{Obermüller, et al.}

\begin{abstract}
	Block-based programming languages like \scratch enable children to be
creative while learning to program. Even though the block-based approach simplifies the creation of programs, learning to program can
nevertheless be challenging. Automated tools such as linters therefore support
learners by providing feedback about potential bugs or code smells in their
programs. Even when this feedback is elaborate and constructive, it still represents purely negative criticism and by construction ignores what learners have done correctly in their programs.
	In this paper we introduce an orthogonal approach to linting: We complement the criticism produced by a linter with positive feedback. We introduce the concept of \emph{code perfumes} as the counterpart to code smells, indicating the correct application of programming practices considered to be good.
	By analysing not only what learners did wrong but also what they did \emph{right} we hope to encourage learners, to provide teachers and students a better understanding of learners' progress, and to support the adoption of automated feedback tools.
	Using a catalogue of 25 code perfumes for \scratch, we empirically demonstrate that these represent frequent practices in \scratch, and we find that better programs indeed contain more code perfumes.
\end{abstract}



\newcommand{\numprojectsBugpattern}{52,783\xspace}
\newcommand{\numBugpattern}{577,206\xspace}
\newcommand{\numprojectsBugpatternwithoutLoose}{50,088\xspace}
\newcommand{\numBugpatternwithoutLoose}{535,420\xspace}
\newcommand{\pearsonPassedPatterns}{-0.16\xspace}
\newcommand{\pvaluePassedPatterns}{0.344\xspace}
\newcommand{\pearsonPassedPatternsBlocks}{-0.408\xspace}
\newcommand{\pvaluePassedPatternsBlocks}{0.012\xspace}
\newcommand{\pearsonPatternBlockCount}{0.007\xspace}
\newcommand{\pvaluePatternBlockCount}{0.969\xspace}
\newcommand{\pearsonPointsPatterns}{-0.055\xspace}
\newcommand{\pvaluePointsPatterns}{0.745\xspace}
\newcommand{\pearsonPointsPatternsBlocks}{-0.265\xspace}
\newcommand{\pvaluePointsPatternsBlocks}{0.109\xspace}
\newcommand{\numprojectsSmell}{68,897\xspace}
\newcommand{\numSmell}{3,357,122\xspace}
\newcommand{\numprojectsSmellwithoutLoose}{68,884\xspace}
\newcommand{\numSmellwithoutLoose}{3,200,465\xspace}
\newcommand{\pearsonPassedSmells}{0.109\xspace}
\newcommand{\pvaluePassedSmells}{0.522\xspace}
\newcommand{\pearsonPassedSmellsBlocks}{-0.353\xspace}
\newcommand{\pvaluePassedSmellsBlocks}{0.032\xspace}
\newcommand{\pearsonSmellBlockCount}{0.249\xspace}
\newcommand{\pvalueSmellBlockCount}{0.137\xspace}
\newcommand{\pearsonPointsSmells}{-0.071\xspace}
\newcommand{\pvaluePointsSmells}{0.673\xspace}
\newcommand{\pearsonPointsSmellsBlocks}{-0.618\xspace}
\newcommand{\pvaluePointsSmellsBlocks}{<0.001\xspace}
\newcommand{\numprojectsPerfume}{73,787\xspace}
\newcommand{\numPerfume}{4,712,055\xspace}
\newcommand{\numprojectsPerfumewithoutLoose}{58,965\xspace}
\newcommand{\numPerfumewithoutLoose}{1,879,518\xspace}
\newcommand{\pearsonPassedPerfume}{0.696\xspace}
\newcommand{\pvaluePassedPerfume}{<0.001\xspace}
\newcommand{\pearsonPassedPerfumeBlocks}{0.469\xspace}
\newcommand{\pvaluePassedPerfumeBlocks}{0.003\xspace}
\newcommand{\pearsonPerfumeBlockCount}{0.889\xspace}
\newcommand{\pvaluePerfumeBlockCount}{<0.001\xspace}
\newcommand{\pearsonPointsPerfume}{0.684\xspace}
\newcommand{\pvaluePointsPerfume}{<0.001\xspace}
\newcommand{\pearsonPointsPerfumeBlocks}{0.443\xspace}
\newcommand{\pvaluePointsPerfumeBlocks}{0.005\xspace}
\newcommand{\averageWMC}{47.53\xspace}
\newcommand{\pearsonPassedBlocks}{0.637\xspace}
\newcommand{\pvaluePassedBlocks}{<0.001\xspace}
\newcommand{\multicolersatz}[2]{
	\mbox{\hbox to 0.5\columnwidth{\phantom{xxxx}$\bullet$ #1\hfill}$\bullet$ #2}
}

\newcommand{\litterbox}{\textsc{LitterBox}\xspace}
\newcommand{\scratch}{\textsc{Scratch}\xspace}
\newcommand{\drscratch}{\textsc{Dr. Scratch}\xspace}
\newcommand{\hairball}{\textsc{Hairball}\xspace}
\newcommand{\qualityhound}{\textsc{Quality Hound}\xspace}
\newcommand{\findbugs}{\textsc{FindBugs}\xspace}
\newcommand{\catnip}{\textsc{Catnip}\xspace}
\newcommand{\whisker}{\textsc{Whisker}\xspace}
\newcommand{\bastet}{\textsc{Bastet}\xspace}
\newcommand{\itch}{\textsc{Itch}\xspace}

\newcommand{\numprojects}{74,907\xspace}

\maketitle

\section{Introduction}\label{sec:intro}

\scratch is a block-based programming language for novice programmers, especially designed to meet the needs and interests of children and young students. The use of visual blocks, the single-window user interface layout,  and the minimal command set reduce complexity and help to overcome initial difficulties like language syntax learning~\cite{maloney2010}. 
While these features and the community helped \scratch to become widely used amongst programming novices and teachers \cite{mcgill2020}, the reduced complexity can neither ensure correctness nor good code quality~\cite{fraedrich2020, hermans2016a, techapalokul2017a}. It can be challenging for young learners to detect and correct bugs and code smells in their programs, and the resulting bad practices can negatively affect their coding habits and computational thinking skills~\cite{hermans2016b}.

To address this problem, program analysis tools such as \litterbox~\cite{Litterbox}, \hairball\cite{boe2013} and \qualityhound\cite{techapaloku2017b} can automatically detect bug patterns and code smells. By identifying and pointing out bugs and code smells in learners' programs such tools can provide support for young programmers during their learning process. When the feedback is provided with explanations and hints it may also help to avoid the same problems in the future. Analysis tools are also helpful to support teachers in analysing their students' current skills and showing deficits that have to be tackled in future lessons. Overall, however, the automated feedback tools entirely rely on pointing out negative aspects of  programs.

While corrective feedback is proven to be very useful in terms of acquiring further cognitive skills~\cite{wisniewski2020power}, \emph{positive} feedback is considered to have better effects on motivational aspects than negative feedback---especially on the task level \cite{hattie2008visible}. 
In contrast, purely negative feedback may harm self-efficacy and autonomy and therefore decrease intrinsic motivation~\cite{wisniewski2020power,ryan2000intrinsic}. A decreased intrinsic motivation in turn might affect dealing with feedback in a negative way: Learners process feedback better when they are motivated and positive feedback activates and motivates learners more than negative feedback~\cite{feedbackLearning}. Therefore, effective learning should include informative feedback about errors as well as correct behaviour to address both cognitive and motivational aspects.

\begin{figure}[tb]
	\centering
	\subfloat[\label{fig:example-perfume}``Perfumed'' code.]{\includegraphics[scale=0.6]{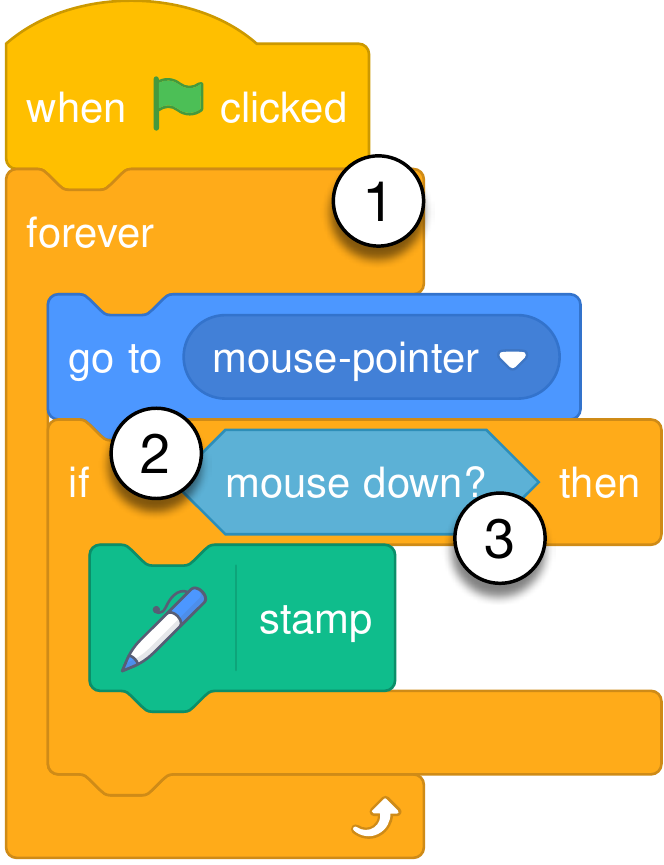}}\hfill
	\subfloat[\label{fig:example-bug}Incomplete code.]{\includegraphics[scale=0.6]{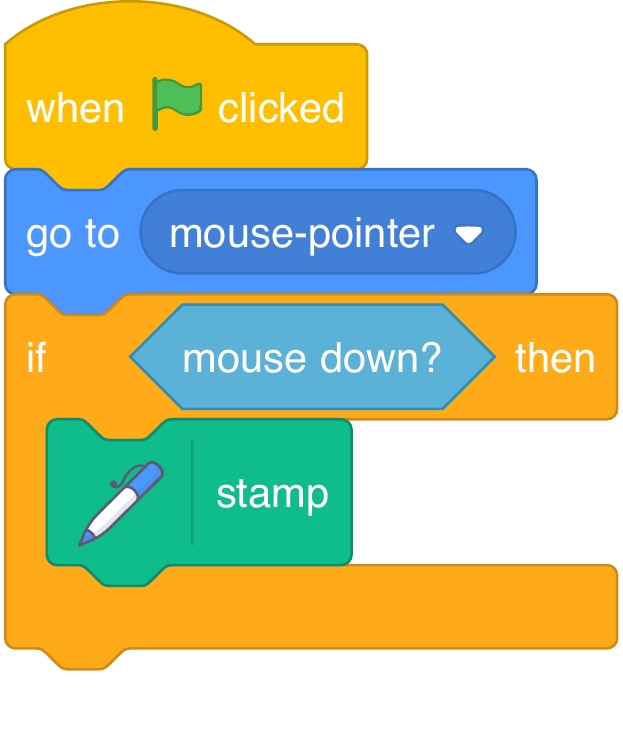}}
	\caption{\label{fig:example_good_practice} Good code practice in a \scratch project: The script continuously \ding{192} checks \ding{193} the mouse down status \ding{194}. A frequent mistake it to omit the loop \ding{192}, in which case the code will not work.}
\end{figure}

In this paper, we introduce the idea of \emph{code perfumes} as a counterpiece to code smells to enable automated program analysis tools to also provide positive feedback. Code perfumes are patterns of code that are considered as good programming practices. For example, consider  Figure~\ref{fig:example_good_practice}: The version of the script in Figure~\ref{fig:example-perfume} continuously \ding{192} checks \ding{193} the mouse down status \ding{194}, which is a common and good pattern of handling repeated events---it is a code perfume. Figure~\ref{fig:example-bug} shows a common failed attempt at achieving the same which lacks the code perfume: Since there is no \controlblock{forever} loop, the check is executed only once, instead of continuously.
Code perfumes make it possible to allow automated program analysis tools to  inform the user not only about bug patterns and code smells in the given program, but also about implementations of good ideas and good coding practices occurring in it. 

\begin{figure}[tb]
	\centering
	\includegraphics[width=\columnwidth]{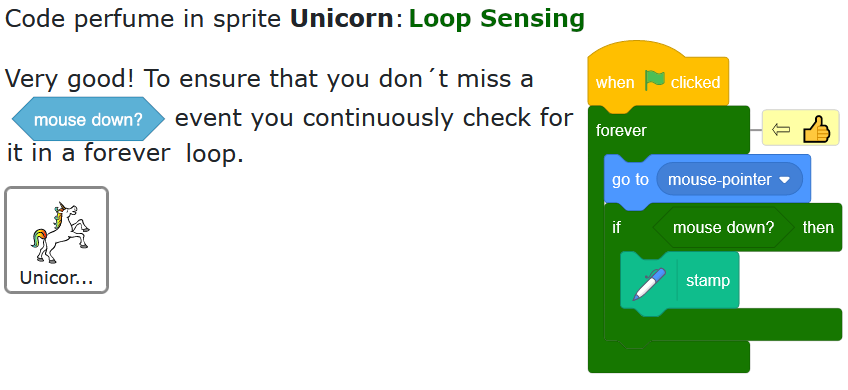}
	\caption{\label{fig:example_feedback}Positive feedback given for correctly implementing a continuous check for an event.}
	\vspace{-1em}
\end{figure}

Figure~\ref{fig:example_feedback} shows feedback commending the learner for applying a code perfume and correctly using the check continuously in a loop.
This potentially has several beneficial effects: The
positive feedback serves as encouragement for young learners and helps teachers
in providing better individual feedback. Furthermore, given information about
positive aspects of student solutions, teachers may be able to ask
contextualised questions that address individual needs and potentials alike,
much more than just knowing what their students did wrong. Finally, learners may be more willing to accept the help from automated feedback tools if these do not just criticise.

In detail, the contributions of this paper are as follows:
\begin{itemize}
\item We introduce the notion of code perfumes as a means to complement bug and code smell finders in automated program analysis tools.
\item We describe and implement a catalogue of 25 code perfumes, covering a wide range of aspects of \scratch programs.
\item We empirically evaluate how frequently these code perfumes occur in practice by analysing a dataset of \numprojects publicly shared \scratch programs.
\item We empirically evaluate how the occurrence of code perfumes relates to correctness of the programs.
\end{itemize}

Our investigation shows that the proposed code perfumes occur frequently in practice, and we find that the number of code perfumes in a program is correlated to how correct the program is. These encouraging findings suggest that code perfumes are a practical means to complement the feedback on errors that existing automated program analysis tools can provide to learners with positive feedback, and to provide teachers with insights into the learning progress of their students.


\section{Background}\label{sec:background}

\subsection{Analysing \scratch Programs}
\scratch\cite{maloney2010} is a block-based programming language that aims to make programming more accessible for novices. The use of different block shapes  prevents syntactical errors, but the resulting code can nevertheless contain problems in terms of code smells or bugs.
Code smells are idioms that decrease the understandability of a program and increase the likelihood of introducing bugs during future modifications~\cite{fowler1999}. Different code smells for \scratch have either been derived from other programming languages \cite{hermans2016a} or have been
defined specially for it~\cite{moreno2014}.
Prior research has demonstrated that code smells have negative effects on
learning~\cite{hermans2016b} as smelly code hampered the learners ability to modify it. Code smells also lower the likelihood of a project being remixed and thus hurt the \scratch community~\cite{techapalokul2017a}. 
In order to find code smells in \scratch projects there are automated tools to analyse the programs like \hairball~\cite{boe2013},  \qualityhound~\cite{techapaloku2017b}, and \litterbox~\cite{Litterbox}.

Bugs, on the other hand, refer to code that does not correctly implement the functionality it is supposed to. While the process of finding and fixing bugs is inherent to coding and learning to program, tools can offer support to learners as well as teachers. Automated tests and program verification are a common means to identify bugs in programs, and have also been introduced to the \scratch domain with tools like \whisker~\cite{stahlbauer2019testing}, \itch~\cite{johnson2016itch}, or \bastet~\cite{stahlbauer2020}. As testing as well as verification require some form of specification, a more accessible approach to support learners is through static code analysis tools (i.e., linters), which can recognise common errors based on rules and patterns~\cite{lintersGeneral, codeAnalysisTools}. 
In particular, misconceptions of learners often manifest in bugs that follow recurring bug patterns which tools like \litterbox~\cite{Litterbox} can automatically detect: A bug pattern is a code idiom that is likely to be a defect~\cite{hovemeyer2004}, and bug patterns for \scratch have been demonstrated to occur frequently~\cite{fraedrich2020}.

\subsection{Feedback and Learning}

Feedback generated by program analysis tools is generally assumed to be effective, in particular because the feedback is returned regardless of other characteristics of the student and is accordingly perceived as less threatening~\cite{hattie2008visible}. 
Computer-based feedback mainly focuses on bad coding practices. Such negative
feedback may be useful in programming education in many ways: Teachers face the
challenge of not only having to assess, but also having to support students with their individual problems when
programming~\cite{michaeli2019current,yadav2016expanding,sentance2017computing},
 and tools can be of help in analysing their students' misunderstandings
and currently lacking skills. Learners, especially more experienced
programmers, mainly seek negative feedback to achieve their
goals~\cite{posAndNegFeedbackGoal}, as corrective feedback helps to acquire
further knowledge and cognitive skills \cite{wisniewski2020power}.

The effects of feedback on intrinsic motivation are generally assumed to be
small, but depend on the type of feedback~\cite{wisniewski2020power}. In
particular, negative feedback might reduce the perceived autonomy and
self-efficacy that in turn both influence intrinsic
motivation~\cite{wisniewski2020power,ryan2000intrinsic}. Positive feedback on
the other hand, tends to lead to a higher intrinsic motivation than negative
feedback~\cite{hattie2008visible}. To ensure these effects, feedback should
support the feeling of autonomy and learners feeling responsible for their
competent performance \cite{ryan2000rewards}. Such authentic positive feedback
increases people's confidence, leading them to expect successful goal
attainment \cite{posAndNegFeedbackGoal}. Accordingly, praise and tangible
rewards do not increase intrinsic motivation~\cite{deci1999meta}. Especially
for learners and novices, positive feedback is effective to keep up motivation
\cite{posAndNegFeedbackGoal}. There is also evidence that more motivated
learners process feedback better~\cite{feedbackLearning}, which means that
positive feedback increasing motivation may enhance the processing of negative
feedback.

Importantly, however, positive feedback is different from praise: Praise
contains little to no content-related information and is therefore among the
least effective types of feedback~\cite{hattie2008visible}. Furthermore, praise
does not answer the three questions of effective
feedback~\cite{hattie2014using}: It does not provide any answers concerning the
intended goals, the previous progress or the future direction. Elaborated
feedback, on the other hand, deals with knowledge about task constraints,
mistakes, processing the task and meta-cognition~\cite{narciss2013designing}.
Positive feedback can support learners of programming only if it answers the
three questions of effective feedback and contains elaborated feedback:
Learners can find out which of the intended goals they attained and which
practices and patterns they can maintain and expand further. While praise can
nevertheless be helpful for motivating students, the resulting motivation will
be extrinsic, and the motivation will decrease when external motivators are
omitted: Montessori \cite{montessori1959absorbent} explains that purely
judgemental comments such as praise do not support children; instead they only
point out what the child might already have known (such as being good or bad at
something), make the child dependent on the teacher and thus also decrease
intrinsic interest.

Despite the potential effects of positive feedback, 
so far, there are only few implementations in automated analysis tools for \scratch programs. \drscratch~\cite{moreno2015} displays praise in form of a short text such as ``You're doing a great job. Keep it up!!!'' and returns points on computational thinking concepts\footnote{http://www.drscratch.org/, last accessed 28.05.21}. 

\section{Code Perfumes}\label{sec:perfumes}

\subsection{What is a Code Perfume?}

Source code that contains quality problems is typically considered to be ``smelly''. A code smell is not necessarily a bug per se, but it may easily lead to bugs after further modifications.

\begin{definition}[Code Smell] A code smell is a code idiom that increases the likelihood of bugs in a program~\cite{fowler1999,fraedrich2020}.
\end{definition}

Based on the metaphor of smelly code, we consider code that is good to also smell good---it is \emph{perfumed}. Therefore, we introduce \emph{code perfumes} as the counterpart to code smells.
%
%
%
%
%
We define code perfumes as follows:

\begin{definition}[Code Perfume]
A code perfume is a code idiom that is indicative of the correct application of a programming concept or pattern.
\end{definition}

Just like with code smells, the question whether a program implements the desired functionality is orthogonal---one can implement the wrong functionality using correctly applied programming concepts, and one can implement the right functionality using smelly code.
The number of code perfume instances can furthermore be easily increased by using many good code parts that do not contribute to the functionality, so code perfumes are not intended to be a single device for automated grading of student solutions.

\subsection{Sources of Code Perfumes}

\label{sec:perfume_sources}

Desirable programming concepts and patterns will differ for programming paradigms and programming languages. In this paper, we focus on perfumes that are suitable for the block based, event driven nature of \scratch.
To define concrete coding patterns that qualify as code perfumes a number of different approaches are presented in this subsection.

\subsubsection{Solution Patterns:}

\begin{figure*}[t]
	\centering
	\subfloat[\label{fig:example-shared}Shared code.]{\includegraphics[width=0.27\textwidth]{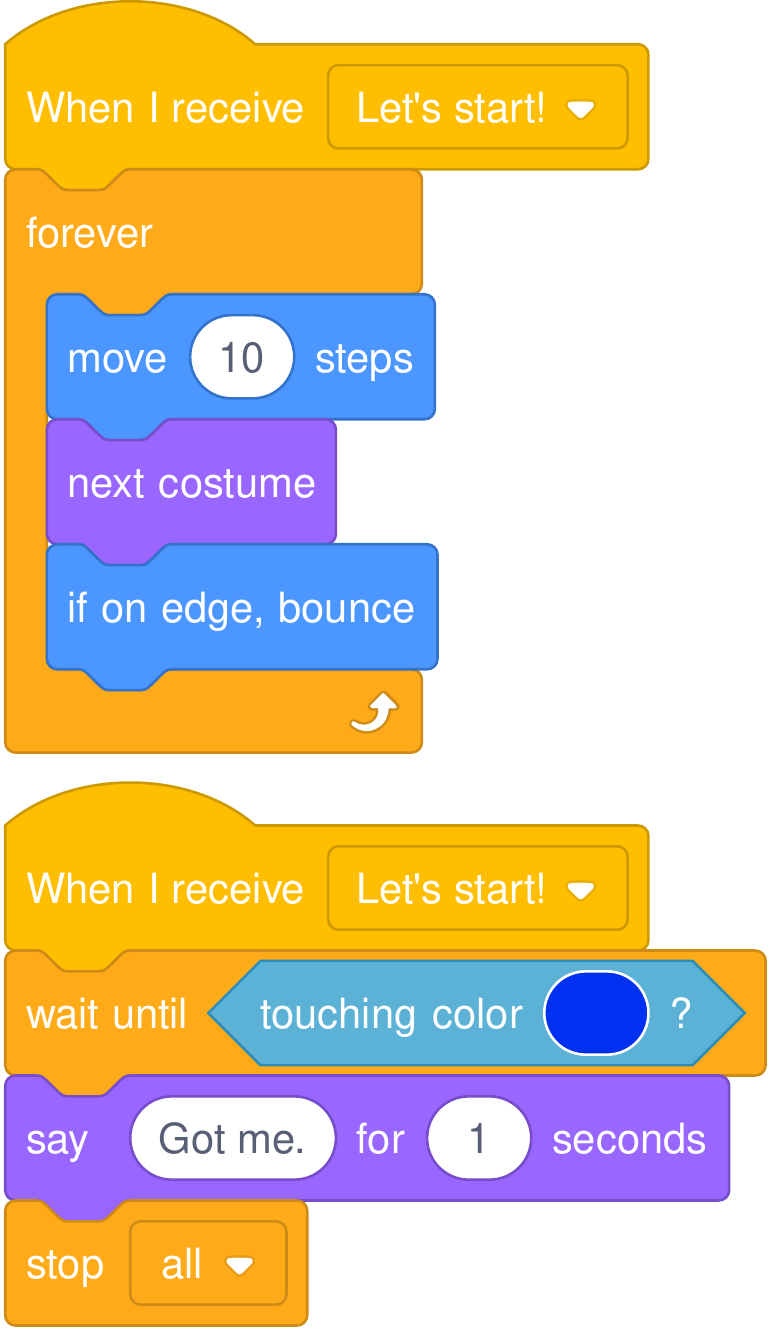}}\hfill
	\subfloat[\label{fig:example-bugpattern}Buggy Version.]{\includegraphics[width=0.34\textwidth]{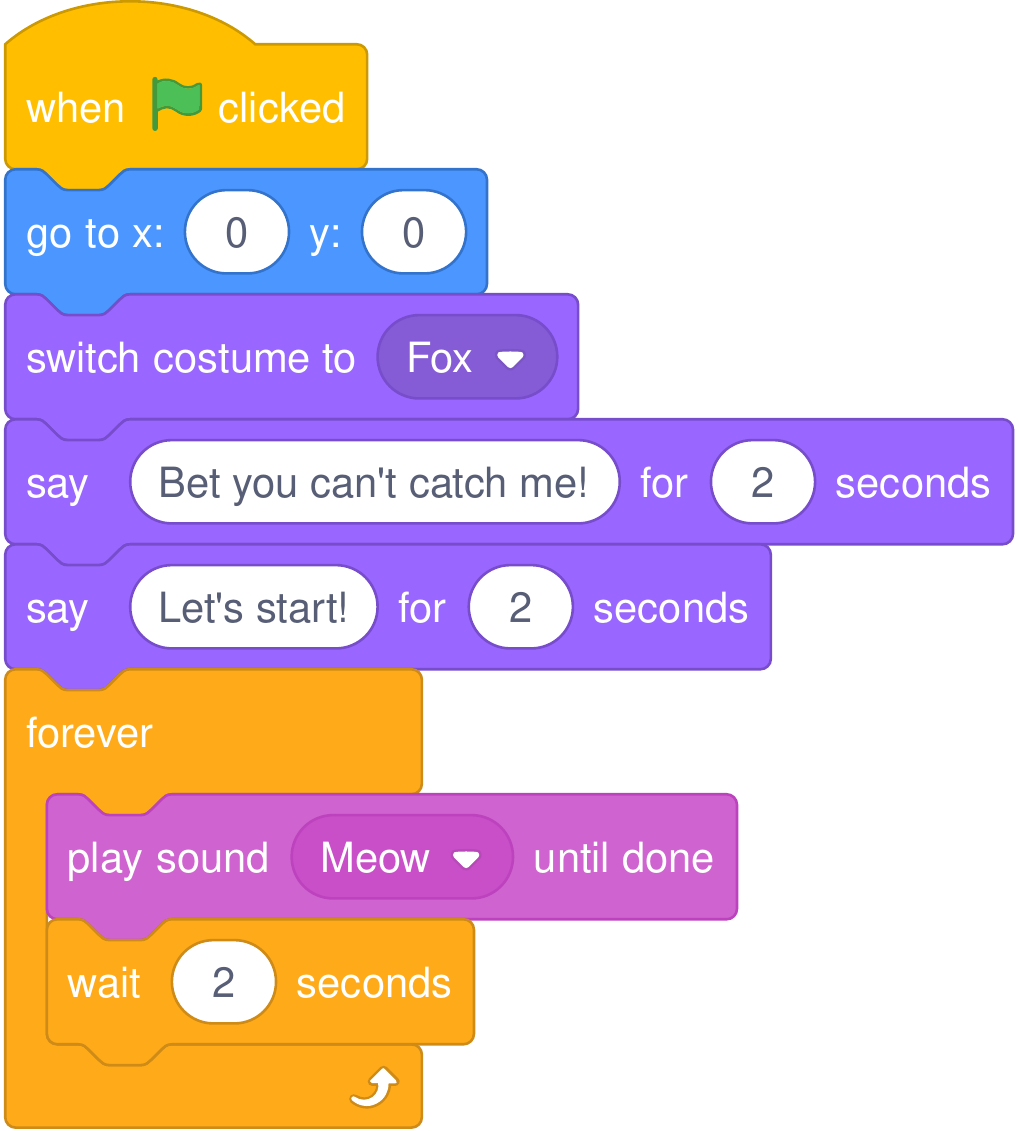}}\hfill
	\subfloat[\label{fig:example-perfume_fix}Perfumed version.]{\includegraphics[width=0.34\textwidth]{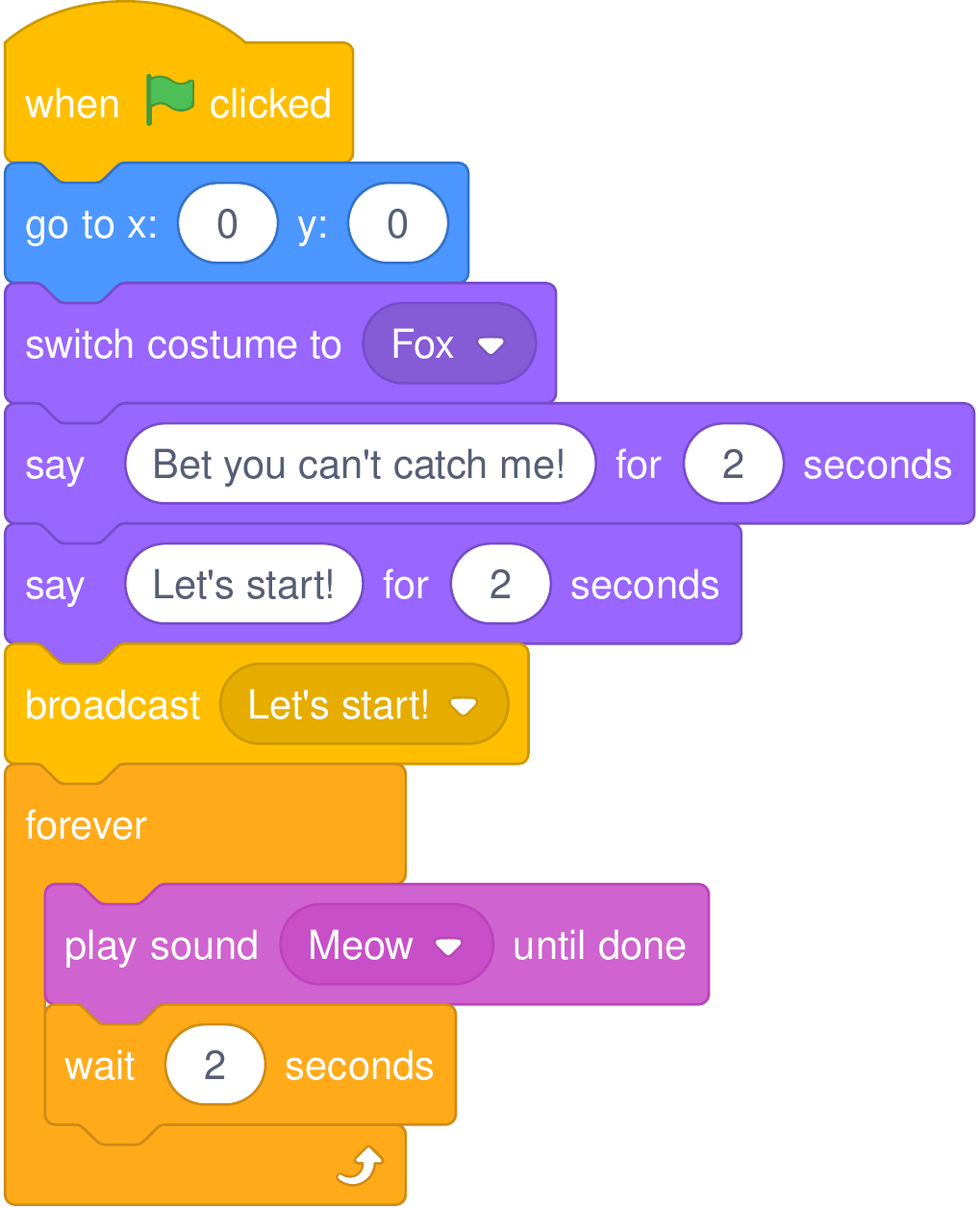}}
	\vspace{-0.5em}
	\caption{Script showing \textit{Message Never Sent} bug pattern that can not start the shared code compared to a script with the corresponding \textit{Correct Broadcast} code perfume with the correct behaviour.}
	\vspace{-1em}
\end{figure*}

Bug patterns in \scratch are defined as compositions of blocks typical of defective code, or common erroneous deviations of a correct code idiom~\cite{fraedrich2020}. When bug patterns are defined as deviations of  correct code idioms, then these underlying correct code idioms represent correct ways of implementing certain behaviour, and can thus be considered perfumed.
As an example consider the bug pattern \textit{Message Never Sent} depicted in Figures~\ref{fig:example-shared} and \ref{fig:example-bugpattern}: The two scripts with the \hatblock{When I receive Let's start!} event handler will never be executed by the buggy version (Figure~\ref{fig:example-bugpattern}), as there is no \controlblock{broadcast Let's start!} block. The use of a broadcast-block in this scenario (Figure~\ref{fig:example-perfume_fix}) represents a correct solution pattern, which in turn provides an opportunity to identify code perfumes. In a similar manner, solution patterns can be defined as counterpieces of many bug patterns.

However, not all bug patterns imply solution patterns. For example, the \textit{Terminated Loop} bug pattern~\cite{Litterbox} checks if there are loops that stop during the first iteration due to a \controlblock{stop} block without guarding condition. The absence of a \controlblock{stop} block in a loop would already be sufficient to avoid the bug pattern, but it hardly justifies positive feedback, as that would simply commend the absence of bugs and not the presence of a solution pattern.

For those cases where we can match a bug pattern with a solution pattern there is the potential to further improve feedback. In particular, if a project contains both the bug pattern and the corresponding solution pattern, then negative and positive feedback could even be linked together, for example:  \emph{``There is a bug here, but you already know how to do that correctly: ... ''}, so that users can learn from their own good code. 

In order to discern perfumes that were inspired by this approach, we annotate corresponding descriptions with a~\solutionpattern symbol.

\subsubsection{Elementary Patterns:}

Amanullah and Bell~\cite{elementarypatterns} suggest that already the use of certain language features implies an understanding of control flow structures and indicates problem solving and logical thinking skills~\cite{elementarypatterns}. For evaluating \scratch programs they therefore provide a list of various command blocks and features, and so called \emph{elementary patterns}, that can indicate such understanding. They provide possible solutions for a number of common (\scratch specific) problems that can help students avoid bad programming style and show a range of good coding techniques. The list of elementary patterns consists of two categories: Loop- and Selection Patterns. Loop patterns, such as the ``Process All Items'' pattern, mainly work on a collection, which is provided as a list in \scratch. Selection patterns refer to conditional constructs with (nested) \controlblock{if then} and \controlblock{if else} blocks. Examples are the ``Whether Or Not'' pattern, consisting of only an \controlblock{if then} statement or the ``Independent Choice'' pattern that contains nested \controlblock{if then} statements when one action depends on several, independent factors.


While these features and patterns have been found
empirically~\cite{elementarypatterns} to be important for developing logical
and computational thinking, they are extremely
under-used~\cite{elementarypatterns}. Indeed code does not have to be
smell free to be worth positive feedback, and simply the correct usage of
specific code structures and concepts can be used for positive feedback as it
indicates development of computational skills.

Code perfumes originating from this category are annotated with a~\elementarypattern symbol.

\subsubsection{Evidence Variables:}

The Progression of Early Computational Thinking (PECT) model~\cite{computationalThinking} aims to find evidence for computational thinking skills in \scratch programs by measuring the presence and level of concrete computational aspects, which are essentially more specific versions of the aforementioned elementary patterns~\cite{elementarypatterns} in a program, through the use of so called \emph{Evidence Variables}.  These variables roughly represent the very basic components of programming and are based on the \scratch block categories. The model focuses on the traditional story-telling and gaming perspective of \scratch, thus Seiter and Foreman~\cite{computationalThinking} propose the following set of Evidence Variables: 

\vspace{0.4em}

\multicolersatz{Looks}{Conditionals}

\multicolersatz{Sound}{Coordination}

\multicolersatz{Motion}{User Interface Event}

\multicolersatz{Variables}{Boolean Expression}

\multicolersatz{Sequence \& Looping}{Operators}


\multicolersatz{Initialise Location}{Initialise Looks}

\mbox{\hbox to 0.6\columnwidth{\phantom{xxxx}$\bullet$ Parallelisation\hfill}}



The first ten Evidence Variables of this list are assigned values of 0
(absent), 1 (basic), 2 (developing) and 3 (proficient), based on the presence
and level of proficiency of the respective construct in a user program. For
example, for \emph{Sequence \& Looping} the levels 1 - 3 are defined as
follows: 1 = Sequence, 2 = Repeat, Forever, 3 = Forever If, Repeat
Until~\cite{computationalThinking}. This categorisation by levels of
sophistication might not be necessary for a linter in our context: Level 0 or 1
represent absent concepts or basic skills and are not particularly worthy of praise, but developing and proficient (level 2 and level 3) implementations of
Evidence Variables can indicate high computational thinking skills and the
correct use of a programming concept, and might therefore qualify as code
perfumes.
The last three Evidence Variables are Initialise Location, Initialise Looks and Parallelisation. They have binary semantics, and as such directly qualify as code perfumes. 

Code perfumes inspired by this approach are annotated with a~\evidencevariables symbol. 

\subsubsection{Game Programming Patterns:}

One central characteristic feature of \scratch is that it easily allows the creation of animated stories and games of all kinds~\cite{maloney2010}. Introducing programming to children in a more entertaining and playful way can not only lead to higher motivation for children, but can also increase their programming knowledge and computational thinking skills~\cite{modelforMeasurement}. Whether or not common game behaviour is implemented correctly gives indication of a student's learning and engagement in computational thinking. 
The Game Computational Sophistication (GCS) model~\cite{modelforMeasurement}
aims to measure this based on the code constructs (patterns) found in the
source code of an analysed game. As an example, the GCS 2.0
model~\cite{modelforMeasurement} comes with an expanded set of common game
programming patterns such as Key/Mouse Control Movement, User-Usability and
Counter. Many of these patterns also fit into the \scratch context or can at
least be adapted accordingly. Consequently, considering the game design aspect
of \scratch these patterns can be considered good practices and a source of
code perfumes. The resulting code perfumes focus on common implementation
choices for classic game behaviour that can help to improve user experience and
functionality.

The game programming perfumes are annotated with a~\gameprogramming symbol.

\subsection{Code Perfumes for \scratch}

\label{sec:perfume_descriptions}


Based on the patterns discussed in Section~\ref{sec:perfume_sources}, we now introduce an initial collection of code perfumes. We anticipate that this list will grow further over time.

{
\setlength{\parskip}{0.5\baselineskip}%
\setlength{\parindent}{0pt}%

\paragraph{\solutionpattern Backdrop Switch:}
Changing to a backdrop in \scratch games or animated stories should likely induce some state alterations in one or more sprites or backdrops.
This can elegantly be implemented using appropriate \looksblock{switch backdrop to} options together with \hatblock{when backdrop switches to} event handlers to start the desired actions.
\textit{Backdrop Switch} is inspired by a possible fix of the \textit{Missing Backdrop Switch} bug pattern~\cite{fraedrich2020}.  

\paragraph{\evidencevariables Boolean Expression:} The presence of combinations of expressions (i.e., \operatorblock{<}, \operatorblock{=} and \operatorblock{>} blocks) and Boolean operators (i.e., \operatorblock{and}, \operatorblock{or} and \operatorblock{not} blocks) can be indicative of attempts to properly simplify control flow \cite{computationalThinking}. 
Note that only instances without Comparing Literals patterns \cite{fraedrich2020} will be reported. 

\paragraph{\gameprogramming Collision:} Continuous collision checks (sprite touches edge or other sprite) that implicate adapted reactions (e.g., move, change look) are used to implement basic game and animation behaviour in \scratch \cite{programmingPatterns, modelforMeasurement}. 

\paragraph{\evidencevariables Conditional Inside Loop:}
Considered as an advanced code structure \cite{programmingPatterns}, this perfume is checking for loops that contain at least one conditional construct. For example an \controlblock{if else} statement within a \controlblock{repeat until} block.

\paragraph{\evidencevariables Controlled Broadcast Or Stop:} The timing and conditions for when to start other scripts via a broadcast, or when to stop scripts, must be correct for a right program behaviour. So a check for a condition, which must be met before broadcasting or stopping, to control both these actions is useful~\cite{youngProgrammersPatterns}. Furthermore, to ensure correct timing, the block responsible for this must be within a loop.

\paragraph{\evidencevariables Coordination:} 
The existence of a \controlblock{wait until} statement in a \scratch program might be a sign of an effort to adapt the coordination of scripts to changing control flows~\cite{computationalThinking}.  

\paragraph{\solutionpattern Correct Broadcast:} 
Properly implemented message broadcasts should at least consist of matching sending and receiving blocks. 
\textit{Correct Broadcast} is inspired by fixes of the \textit{Message Never Received} and the \textit{Message Never Sent} bug pattern~\cite{fraedrich2020}. 

\paragraph{\solutionpattern Custom Block Usage:} 
To identify solutions of subtasks that might be reusable and to implement appropriate custom procedures is considered to be good programming practice. 
This finder is inspired by fixes of the \textit{Call Without Definition} bug pattern \cite{fraedrich2020}. It detects the presence and use of custom blocks.

\paragraph{\gameprogramming Directed Motion:} 
Controlling sprite movement by keyboard inputs is a common task in games and animated stories. A simple implementation consists of a \hatblock{when key pressed} event handler followed by \motionblock{point in direction} and \motionblock{move steps} statements \cite{modelforMeasurement, programmingPatterns}. 

\paragraph{\gameprogramming Gliding Motion:} 
This finder reports another simple implementation to manipulate sprite movement: a \hatblock{when key pressed} event handler followed by one or more \motionblock{glide secs to} statements~\cite{modelforMeasurement,programmingPatterns}. 

\paragraph{\evidencevariables Initialisation of Looks:} 
Defining the start state of games and animated stories is especially useful, since \scratch does not perform any default resetting of attributes automatically. Furthermore, it is considered a good programming practice to think about desired initial states of program executions.
Look blocks like costume or backdrop setter statements being present in \hatblock{when green flag clicked} scripts are reported by this finder. This might indicate that learners tried to solve a subtask of the \textit{defining a start state} problem. The presence of this and the next pattern is used by Seiter and Foreman~\cite{computationalThinking} to measure computational thinking skills of students.  

\paragraph{\evidencevariables Initialisation of Positions:} 
This perfume finder reports position setter statements being present in \hatblock{when green flag clicked} scripts possibly indicating that learners tried to solve another subtask of the \textit{defining a start state} problem.

\paragraph{\elementarypattern List Usage:} 
The existence of list-statements in \scratch programs might be a sign of an effort to hold and process a number of values efficiently.  

\paragraph{\solutionpattern Loop Sensing:} 
Continuously checking for touch or key events inside a \controlblock{forever} or \controlblock{repeat until} loop is a useful pattern to implement event processing in \scratch.
This perfume is inspired by a possible fix of the bug pattern \textit{Missing Loop Sensing} \cite{fraedrich2020}.

\paragraph{\solutionpattern Matching Parameter:}
Properly implemented custom blocks consist at least of a signature containing a complete parameter list: all parameters, that are used inside a custom block, are to be present in the list.     
This perfume finder is inspired by fixes of the \textit{Orphaned Parameter} bug pattern \cite{fraedrich2020}: It detects whether all parameters used are also declared in the custom block.

\paragraph{\gameprogramming Mouse Follower:} 
Sprite movement can be controlled by mouse input.
This behaviour can be implemented in \scratch by a loop containing either a \motionblock{go to mouse-pointer} statement or a combination of \motionblock{point towards mouse-pointer} and \motionblock{move steps} statements \cite{programmingPatterns}.

\paragraph{\solutionpattern Movement In Loop:} To avoid \textit{Stuttering Movement} \cite{fraedrich2020} when controlling sprites by keyboard input it is recommended to use a loop with a conditional containing a \sensingblock{key pressed?} expression and appropriate actions.

\paragraph{\elementarypattern Nested Conditional Checks:} Nested conditional checks (i.e., nested \controlblock{if then} and \controlblock{if else} blocks) can be seen as advanced code structures~\cite{programmingPatterns, elementarypatterns}.

\paragraph{\solutionpattern Nested Loops:} The presence of nested loops, where the inner one is accompanied by other blocks preceding or following it to not have a \textit{Nested Loop} smell \cite{Litterbox}, might be indicative for attempts to implement advanced control flow \cite{programmingPatterns}. 

\paragraph{\gameprogramming Object Follower:} 
In some games or animations one sprite follows another for at least a certain time \cite{programmingPatterns}. This can be implemented using
a loop containing a \motionblock{point towards} statement targeting the other sprite, followed by a \motionblock{move steps} statement. 

\paragraph{\evidencevariables Parallelisation:}
 The presence of two scripts with the same hat block can be indicative of attempts to implement independent subtasks more clearly and readably \cite{computationalThinking}. 

\paragraph{\gameprogramming Say Sound Synchronisation:} A nice way to enhance interaction between program and player is to use both \looksblock{say} and \soundblock{play sound} blocks in a synchronous way to let sprites talk. However, this say sound synchronisation is not straightforward in \scratch. It can be implemented by placing a \soundblock{play sound file} block, playing a message, right after the \looksblock{say} block that shows the message in a speech bubble. As soon as the sound file ends, the speech bubble must be cleared by using an empty \looksblock{say} block afterwards \cite{boe2013}.

\paragraph{\gameprogramming Timer:} 
Timing durations is a useful subtask in many \scratch programming problems.
This finder reports the usage of a variable that is changed repeatedly (inside, e.g., a \controlblock{forever} loop) by a fixed value in combination with a \controlblock{wait seconds} statement.
\cite{programmingPatterns, modelforMeasurement}.

\paragraph{\solutionpattern Useful Position Check:} 
Checking position and distance values can be quite error-prone since floating point values are used and have to be compared. A bigger-than or less-than operator to compare values can be a fix to the \textit{Position Equals Check} bug pattern~\cite{fraedrich2020}.

\paragraph{\solutionpattern Valid Termination Condition}: The \controlblock{repeat until} statement requires a termination condition, otherwise the loop will run forever and code following the loop will never be executed. This perfume is inspired by a possible fix of the \textit{Missing Termination Condition} bug pattern \cite{fraedrich2020}.

}

\subsection{Analysing Code for Code Perfumes}

\begin{figure}[tb]
  \centering
	\includegraphics[width=0.8\columnwidth]{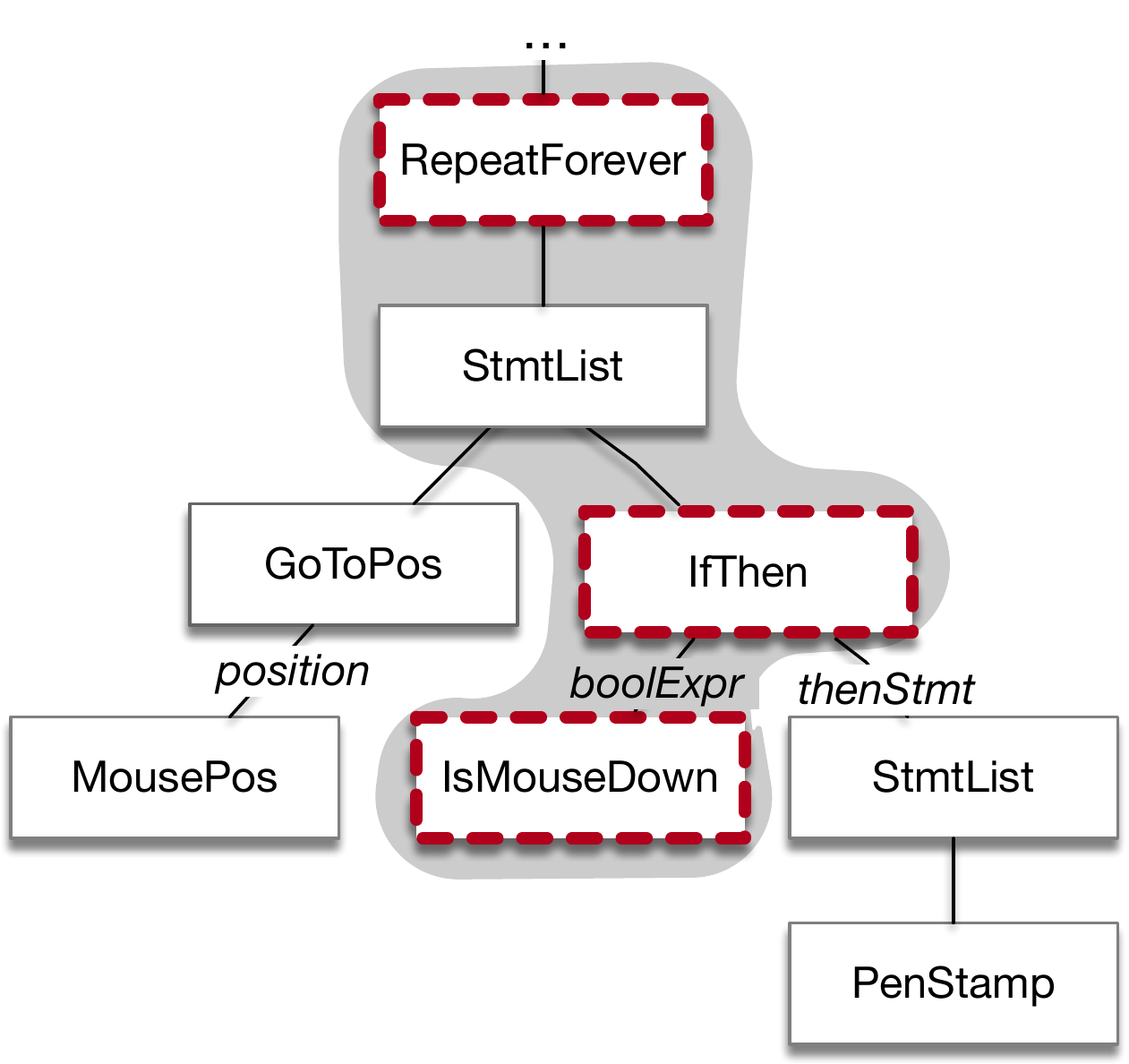}
	\caption{\label{fig:example_ast} Perfume pattern in the abstract syntax tree.}
\end{figure}

The process of finding code perfumes in a given program is similar to how linters typically find issues. First, we build an Abstract Syntax Tree (AST) from the source code, which encodes the structure of the program as well as the type and relation of its elements in an easily traversable tree datastructure. For each component of the \scratch program (i.e., blocks and drop down menus) there is a distinct representation as node in this AST. For example, Figure~\ref{fig:example_ast} shows the AST constructed from the example \scratch program in Figure~\ref{fig:example_good_practice}. 

Each code perfume represents a pattern specifying a combination of nodes in the AST, which together construct a correct implementation of a programming concept. Finding instances of code perfumes therefore is a matter of matching these patterns on the AST. For example, to check whether the AST in Figure~\ref{fig:example_ast} contains an instance of the \emph{Loop Sensing} code perfume, the pattern describes a \controlblock{forever} statement which includes an \controlblock{if then} statement that uses a \emph{sensing} block as Boolean expression. As highlighted in Figure~\ref{fig:example_ast}, the AST contains all of these nodes in the required arrangement, and thus represents an instance of the \emph{Loop Sensing} code perfume.

For each of the code perfumes defined in Section~\ref{sec:perfume_descriptions} we specified a similar pattern of statements and implemented it as an AST-visitor.


\section{Evaluation}\label{sec:evaluation}

To evaluate the concept of code perfumes on \scratch projects, we investigate the
following research questions:
\begin{itemize}
	\item \textbf{RQ1:} How common are code perfumes?
	\item \textbf{RQ2:} Are code perfumes related to correctness?
\end{itemize}

\subsection{Experimental Setup}

\subsubsection{Dataset:}
To answer RQ1 we use the dataset by Frädrich et al.~\cite{fraedrich2020} to
study bug patterns. It consists of \numprojects \scratch projects mined over
the course of 3 weeks. The dataset excludes remixes, which might otherwise skew
results due to duplicated code.

To answer RQ2 we use the 37 Fruit Catching game solutions which were created by children of a sixth and a seventh grade class and used in the evaluation of the \whisker testing tool~\cite{stahlbauer2019testing}. The game contains three sprites: a bowl that the player can move left or right with the cursor keys, an apple and a banana that appear at random positions at the top of the stage and then drop down to the bottom. The goal is to catch as much fruit as possible within a given time, and when missing an apple the game is lost. In addition to the code itself, the dataset includes a set of 28 automated tests~\cite{stahlbauer2019testing}.

\subsubsection{Analysis tool: } 
We implemented the code perfumes listed in Section~\ref{sec:perfume_descriptions} in the open source \litterbox framework. This tool already has the capabilities of statically analysing \scratch projects, and we added new finders for these perfumes following the procedure described by Fraser et al.~\cite{Litterbox}. 

We furthermore use \whisker~\cite{stahlbauer2019testing}, an automatic testing tool for \scratch, to check the solutions of the Fruit Catching game for correctness. \whisker simulates user inputs automatically and then checks if the program reacts as specified in the tests that are executed (e.g., that the bowl moves right when the right arrow key is pressed). For this we use the test suite provided as part of the \whisker study. It contains 28 individual test cases checking different correctness aspects of the task.

\subsubsection{RQ1:}
To answer RQ1 we applied \litterbox to the dataset containing random \scratch projects. For checking how common code perfumes are we consider the total number of code perfumes found, the instances found for each type of perfume, as well as the number of projects containing at least one perfume. Furthermore we consider the average complexity of the projects containing code perfumes measured as weighted method count (i.e., sum of cyclomatic complexities of all scripts).

\subsubsection{RQ2:}
To answer RQ2 we applied \litterbox to the Fruit Catching dataset to obtain the number of code perfumes, bug patterns and code smells in the solutions. We used \whisker to determine the correctness of the programs in terms of the number of tests passed. We then compared code perfumes, bug patterns and smells by their relationship to the number of tests passed.

\subsubsection{Threats to Validity:}
Our experiments are based on the sample of projects mined by Frädrich et al.~\cite{fraedrich2020}, and might not generalise to other data. In particular, only publicly shared projects can be mined, while unfinished, unshared programs might have other properties. Similarly, we only used one task and student solutions from two classes as well as one \whisker test suite for answering the question if code perfumes can indicate correctness, so the results may not generalise to other tasks, classes or test suites. While we analysed how frequent code perfumes are and how they are related to correctness, a study of whether reporting perfumes influences motivation and learning outcomes is a matter of future work.

\subsection{RQ1: How common are code perfumes?}

\begin{table}
\centering
\caption{Number of perfume instances found in total and number of projects containing the perfume.}
\label{tab:perfume_table}
\resizebox{\columnwidth}{!}{%
\begin{tabular}{lrrr}
\toprule
                     Perfume & \# Perfumes & \# Projects &  AVG WMC \\
\midrule
             Backdrop Switch &    140,546 &      8,641 &    74.03 \\
          Boolean Expression &  1,037,703 &     35,073 &    77.18 \\
                   Collision &     59,364 &     14,464 &    72.26 \\
     Conditional Inside Loop &    411,149 &     38,935 &    70.28 \\
Controlled Broadcast Or Stop &     35,566 &     10,205 &    99.66 \\
                Coordination &     66,837 &     10,162 &   131.92 \\
           Correct Broadcast &    297,040 &     37,350 &    70.75 \\
          Custom Block Usage &    174,152 &      9,796 &   114.99 \\
             Directed Motion &      8,039 &      2,116 &    30.01 \\
              Gliding Motion &      7,294 &      1,810 &    43.44 \\
     Initialisation Of Looks &    473,814 &     54,065 &    57.66 \\
  Initialisation Of Position &    131,560 &     38,824 &    62.89 \\
                  List Usage &    130,290 &      4,082 &   172.18 \\
                Loop Sensing &    221,780 &     29,115 &    69.97 \\
          Matching Parameter &     50,721 &      4,674 &   152.39 \\
              Mouse Follower &      5,521 &      4,049 &    79.53 \\
            Movement In Loop &     22,165 &     11,665 &    98.92 \\
   Nested Conditional Checks &    134,673 &     14,256 &   127.23 \\
        Nested Loops Perfume &     30,523 &     11,359 &   116.37 \\
             Object Follower &      9,815 &      3,970 &   146.93 \\
             Parallelisation &  1,142,319 &     70,519 &    49.74 \\
   Say Sound Synchronisation &        155 &         41 &    93.54 \\
                       Timer &     12,955 &      7,627 &   104.38 \\
       Useful Position Check &     28,630 &      7,482 &   126.00 \\
           Valid Termination &     79,444 &     16,447 &   105.23 \\
\midrule
     Total & \numPerfume & \numprojectsPerfume &       93.90\\
\bottomrule
\end{tabular}%
}
\end{table}

We found instances of all 25 code perfumes in the dataset of random projects. In total, there are \numPerfume code perfume instances, and \numprojectsPerfume contained at least one code perfume. Table~\ref{tab:perfume_table} summarises the number of code perfume instances found for each type, the number of projects containing at least one instance of the respective perfume and the average weighted method count of these projects. Note that the individual numbers of projects containing one type of perfume may not add up to the total number of projects containing at least one perfume, as some projects contain more than one type of perfume and are thus counted in more than one category.

Considering the number of projects containing code perfumes, the most common perfume is \textit{Parallelisation} (70,519). Using highly concurrent, parallel scripts is a key factor of the event-driven paradigm followed by \scratch, so the large quantity of \textit{Parallelisation} perfumes found is not surprising. The average weighted method count (WMC) of 49.74 of the projects containing this perfume also suggests that this concept is already used in  smaller projects than other perfumes.

The frequent occurrence of \textit{Initialisation of Looks} (54,065) and \textit{Initialisation of Positions} (38,824) is also intuitive, since initialising the looks and locations of sprites in \scratch is usually necessary for programs to work correctly. Again the average WMC is comparatively low, showing that initialisation is already important at an early state of programming to maintain a correct execution of the program.

The least common perfume is \textit{Say Sound Synchronisation} (41), which is not directly related to programming concepts. It likely is only useful in specific types of animation projects, therefore lower numbers can be expected. 
The infrequent occurrence of the motion related  \textit{Gliding Motion} (1,810), \textit{Directed Motion} (2,116), and \textit{Movement in Loop} (11,665) perfumes matches prior research~\cite{fraedrich2020} which found frequent occurrences of the \textit{Stuttering Movement} bug pattern. It appears that learners prefer to follow the event-driven but simpler approach of handling motion through dedicated event-handling blocks, even though this results in stuttering movement.

The quite low number of only 4,082 projects with \textit{List Usage} code perfume further supports previous research~\cite{elementarypatterns}. The  average WMC of the projects using lists is also the highest one of all code perfumes with 172.18, suggesting that lists are one of the most advanced concepts in \scratch. Similarly, code perfumes related to custom blocks (i.e., \textit{Custom Block Usage} or \textit{Matching Parameters}) are less frequent and usually found in more complex projects, which again matches prior findings on bug patterns~\cite{fraedrich2020}. 

Beside the list and custom block related code perfumes the \textit{Coordination} (131.92) and \textit{Object Follower} (146.93) stand out for the high average WMC of the projects they are contained in. 
The average WMC of \textit{Coordination} may be explained by the fact that the functionality of a \controlblock{wait until} block can also be achieved by a combination of \controlblock{forever} loop and \controlblock{if then} block, two blocks users are probably more confident to use as the high number of \textit{Conditional Inside Loop} perfumes implies (i.e., 38,935 in contrast to only 10,162 projects using \textit{Coordination}). 
This is further supported by the quite low average WMC of \textit{Conditional in Loop} indicating that even simple projects use this paradigm rather than a \textit{Coordination} with a \controlblock{wait until} block.
In the case of \textit{Object Follower} the high complexity may simply be a result of the types of projects in which the perfume can be usefully applied. For comparison, the \textit{Mouse Follower} perfume is conceptually similar, but as it can be used in projects with only a single sprite, the average WMC is only 79.53.
We also note a high average WMC for the \textit{Nested Conditional Checks} and \textit{Nested Loops Perfume}, but these are not surprising as nesting loops and \controlblock{if then}/\controlblock{if else} blocks by definition contribute to an increased complexity as measured by WMC. 

The ``Perfumes''-column of Table~\ref{tab:perfume_table} shows that some of the perfumes also occur multiple times within the same project. Notably, the \textit{Parallelisation} code perfume therefore not only occurs in the most projects, but it also shows most perfume instances (1,142,319) overall. This again is due to the inherently parallel nature of \scratch programs. It has to be noted that a very high amount of parallelisation in a project could also lead to debugging problems if the students are not fully aware how the parallel scripts work \cite{meerbaum2011}, so for evaluating individual students this value should not be taken as the sole grading factor but instead be combined with other characteristics of the project.

The other code perfume standing out due to its high total number of instances found is \textit{Boolean Expression} with 1,037,703. This is also very natural as the Boolean operators in \scratch are a crucial part to regulate the control flow without having to nest multiple \controlblock{if then} blocks. In addition an \operatorblock{and} block is a simple way to prevent a \textit{Nested Loops} smell. A contributing factor to this high number is that our implementation flags each Boolean operator in a nested expression separately, as each represents a correctly used Boolean expression.

There is a much bigger difference between projects using \textit{Initialisation of Looks} (54,065) to total number of perfumes (473,814) than for \textit{Initialisation of Positions} with 38,824 projects containing a total of 131,560 perfumes. This is because the perfume checking for the looks blocks has multiple possible initialisations (e.g., visibility, size and costume) whereas the other one just looks at the position.

The \textit{Mouse Follower} code perfume has the lowest proportion of instances (5,521) to projects containing at least one perfume (4,049). This can be explained with the nature of the pattern: The user wants to control a sprite with the mouse movements, and since a computer normally only has one mouse, controlling multiple sprites is usually not useful.

Lists may be one of the least used features of \scratch~\cite{elementarypatterns}, but users knowing how to handle lists use them frequently, as indicated by the 130,290 instances of \textit{List Usage} in only 4,082 projects having this perfume.
The only perfume not to reach a four digit count is \textit{Say Sound Synchronisation} with only 155 instances found in total. Nonetheless, even this perfume indicates that if a user knows this specific practice it will be used multiple times in one project.

\summary{RQ1}{Code perfumes very frequently appear in \scratch{} programs of varying complexity and can be automatically detected. We found code perfumes in 98\% of our dataset.}

\subsection{RQ2: Are code perfumes related to correctness?}

\begin{figure}[tb]
	\centering
	\subfloat[\label{fig:scatter_perfume_correctness}Passed tests vs. code perfumes.]{\includegraphics[width=0.49\columnwidth]{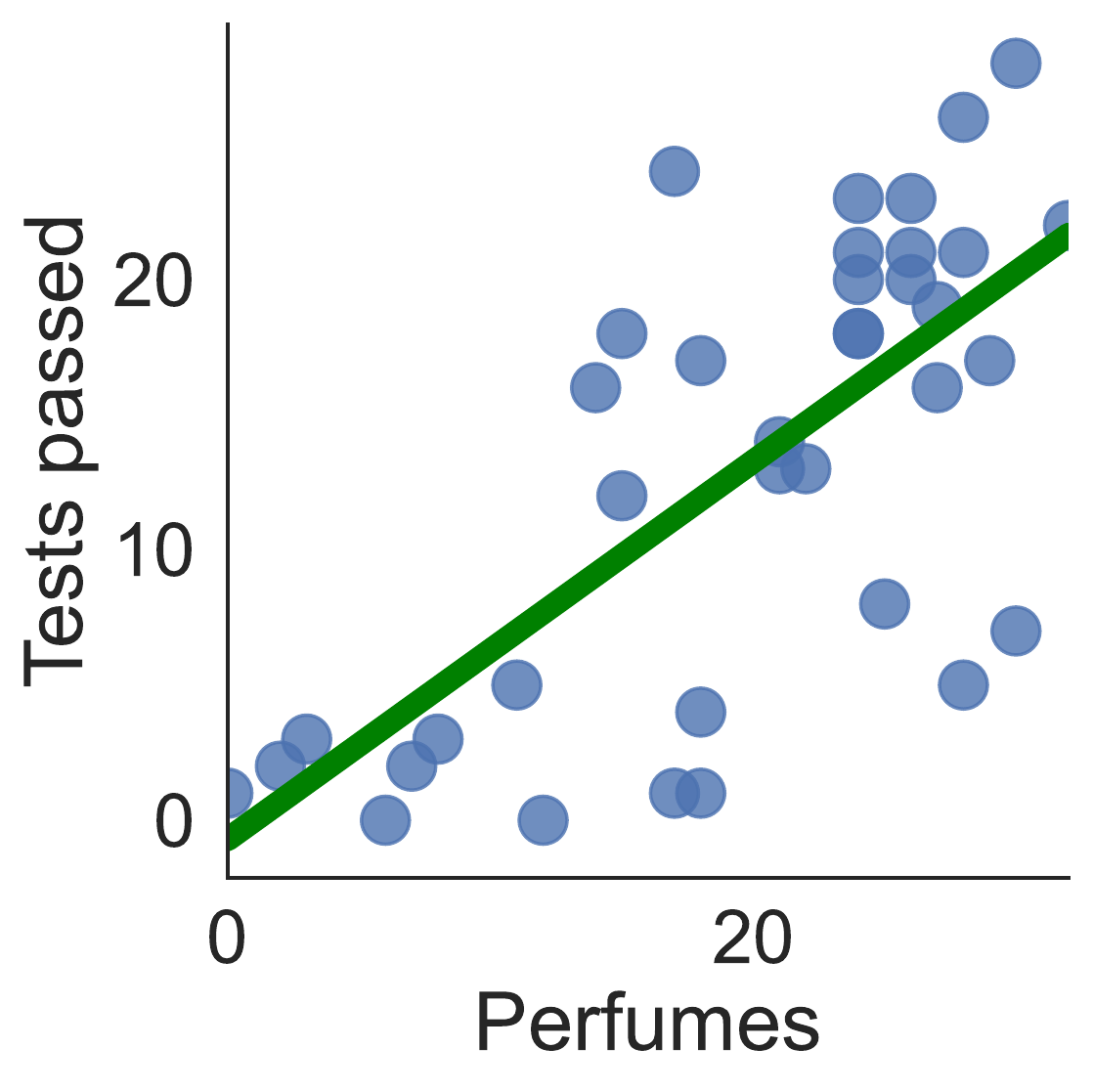}}\hfill
	\subfloat[\label{fig:perfume_blocks_scatter}Code blocks vs code perfumes.]{\includegraphics[width=0.49\columnwidth]{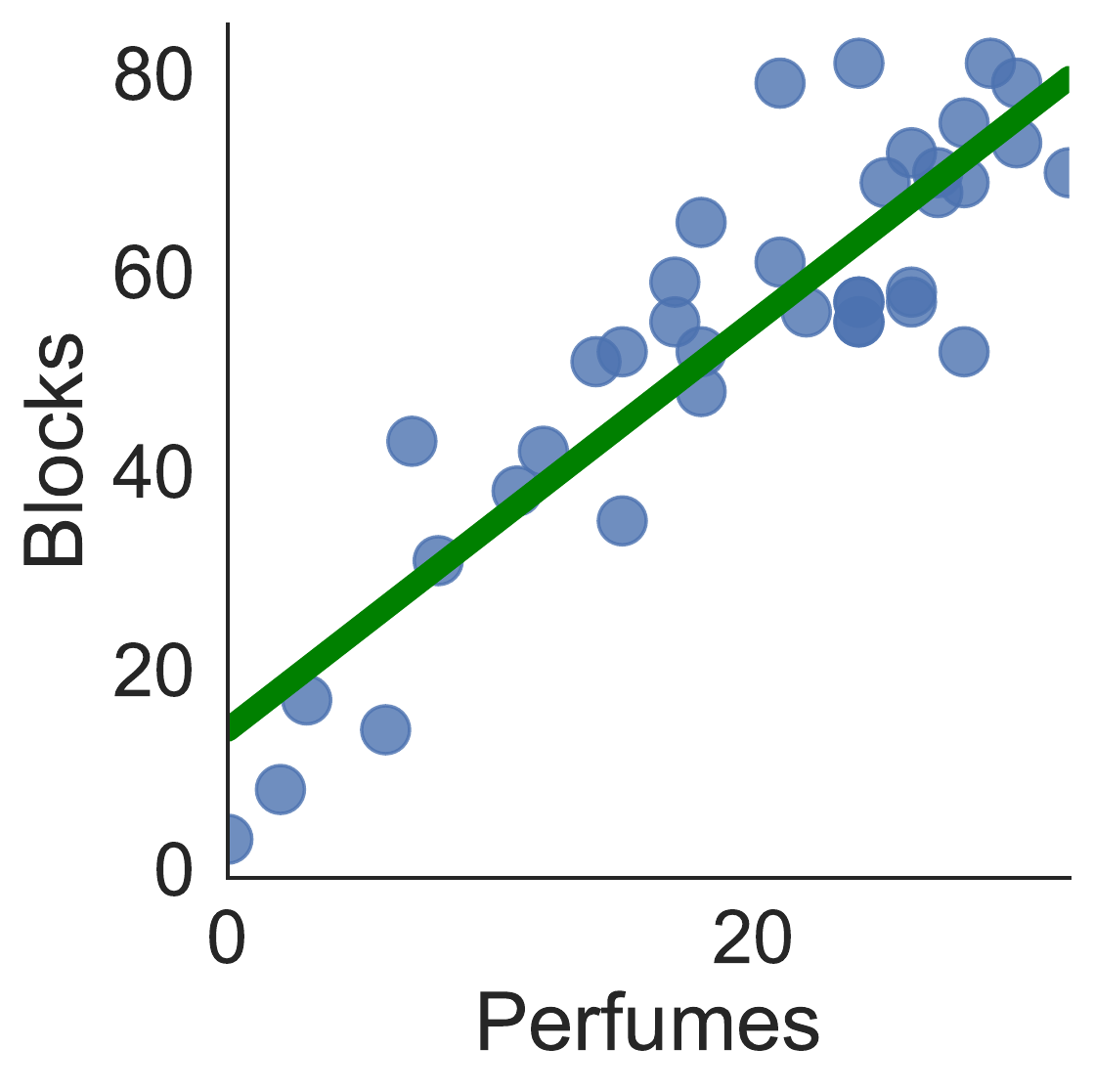}}
	\caption{\label{fig:scatter_plots} Relation between code perfumes and correctness and size metrics.
		}
	\vspace{-0.5em}
\end{figure}

Using the \scratch projects and \whisker tests provided by Stahlbauer et al.~\cite{stahlbauer2019testing}, we can determine for each project how many of the functional requirements that the students were tasked to implement are met, as well as how many code perfumes their projects contain. Since all projects are solution attempts for the same task we hypothesise that, the more tests a project has passed, the more perfumes should be in the code, as more concepts have been used correctly.


The results of the comparison between tests passed and the number of code perfumes found in each project are shown in Figure~\ref{fig:scatter_perfume_correctness}. 
We included a sample solution, which is the only project fulfilling all specifications and thus passing all tests, and this also has the second highest number of code perfumes. We can also see that projects with very few passed tests tend to also only have few code perfumes. 
This results in a Pearson's correlation coefficient between passed tests and number of code perfumes of r=\pearsonPassedPerfume (p\pvaluePassedPerfume). 
That is, there is a moderate correlation between correctness of the projects according to the passed \whisker tests and the number of code perfumes.

There are several outliers that did fairly well in the \whisker tests but do not contain many perfumes, and vice versa. These are solutions that work well but do not exhibit many good coding practices or solutions that have good code quality for scripts that do not contribute to the functionality of the project. 

For example, one solution passed only 5 tests but has 28 perfume instances in its code. The student implemented many functionalities right, which resulted in many code perfumes (e.g., the program stop is controlled inside an \controlblock{if then} block and the countdown is implemented to fit the \textit{Timer} pattern). However, the solution has one crucial failure which results in the low count of passed tests: the Apple is at the bottom of the stage and does not go back to the top when the program starts. Therefore the program ends almost immediately when the green flag is clicked, and only tests related to visibility and the termination of the game pass, even though other functionalities are implemented. In this case the code perfumes indicate that there is good code that the tests do not detect.

Another project contains only 17 code perfumes but passes 24 of the \whisker tests. The code works fine, but it lacks good practices, for example the movements of the bowl are implemented in a way that they exhibit the \textit{Stuttering Movement} bug pattern and not using a \textit{Movement In Loop} perfume. The code is furthermore missing things like an initialisation of the visibility of the sprites, which does not matter for most of the tests. Consequently the code perfumes indicate code that is functional \textit{and} does show good practice.

Figure~\ref{fig:perfume_blocks_scatter} shows the relation between perfumes and number of blocks. There is a strong correlation between these two factors of a project (r=\pearsonPerfumeBlockCount, p\pvaluePerfumeBlockCount). 
Intuitively, when a task is given in detail, a student solution that consists of more code is likely to solve more subtasks and thus might contain more code perfumes.
If students struggle, they likely have to spend more time on solving individual problems and thus produce less code. Considering the correlations with size and with correctness, the number of code perfumes thus can provide some feedback to teachers on whether their students are performing well or are struggling.

\begin{figure}[tb]
	\hfill
		\subfloat[\label{fig:scatter_bugpattern_correctness}Passed tests vs. bug patterns]{\includegraphics[width=\columnwidth/2]{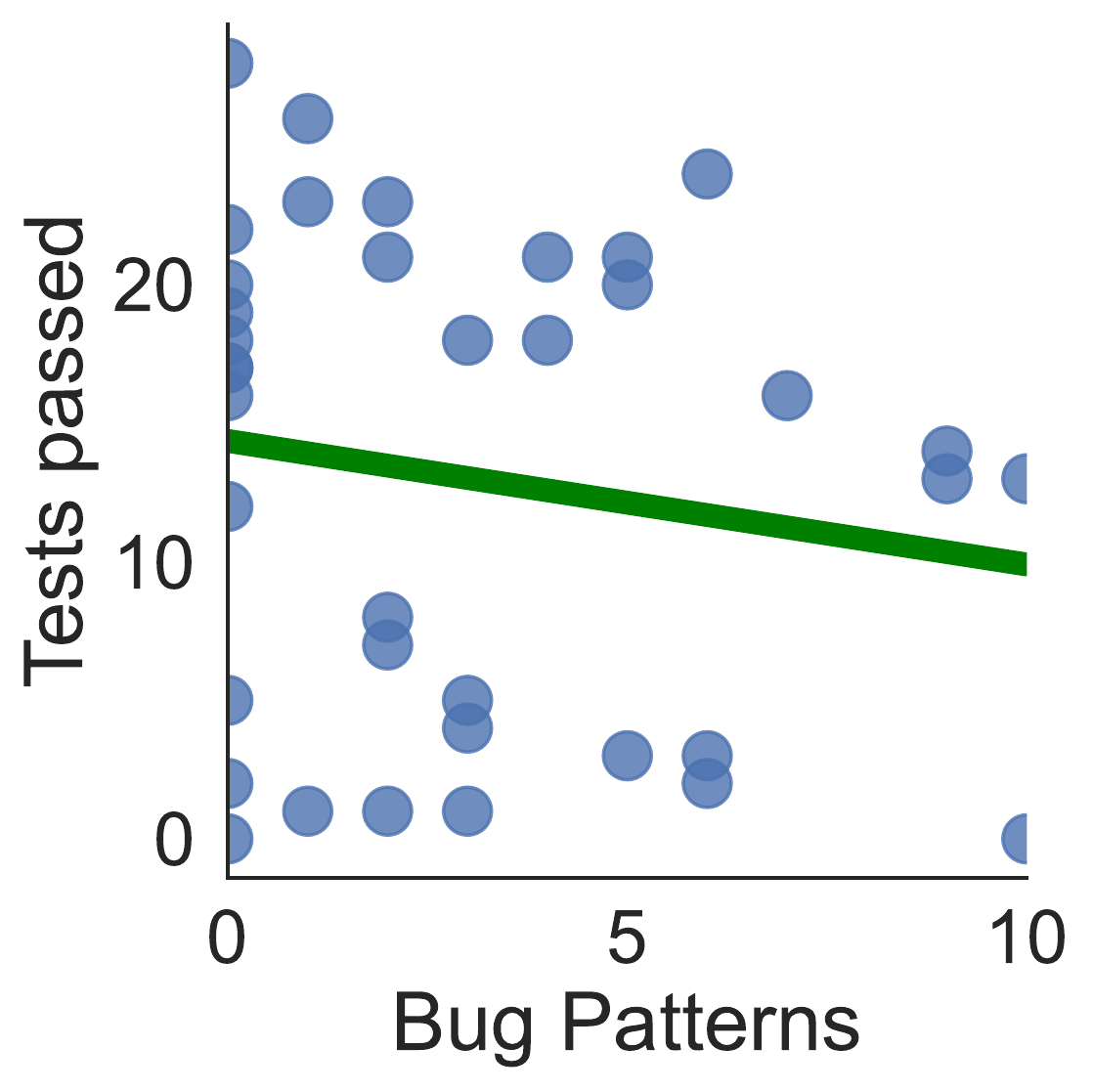}}\hfill
		\subfloat[\label{fig:scatter_smell_correctness}Passed tests vs. code smells]{\includegraphics[width=\columnwidth/2]{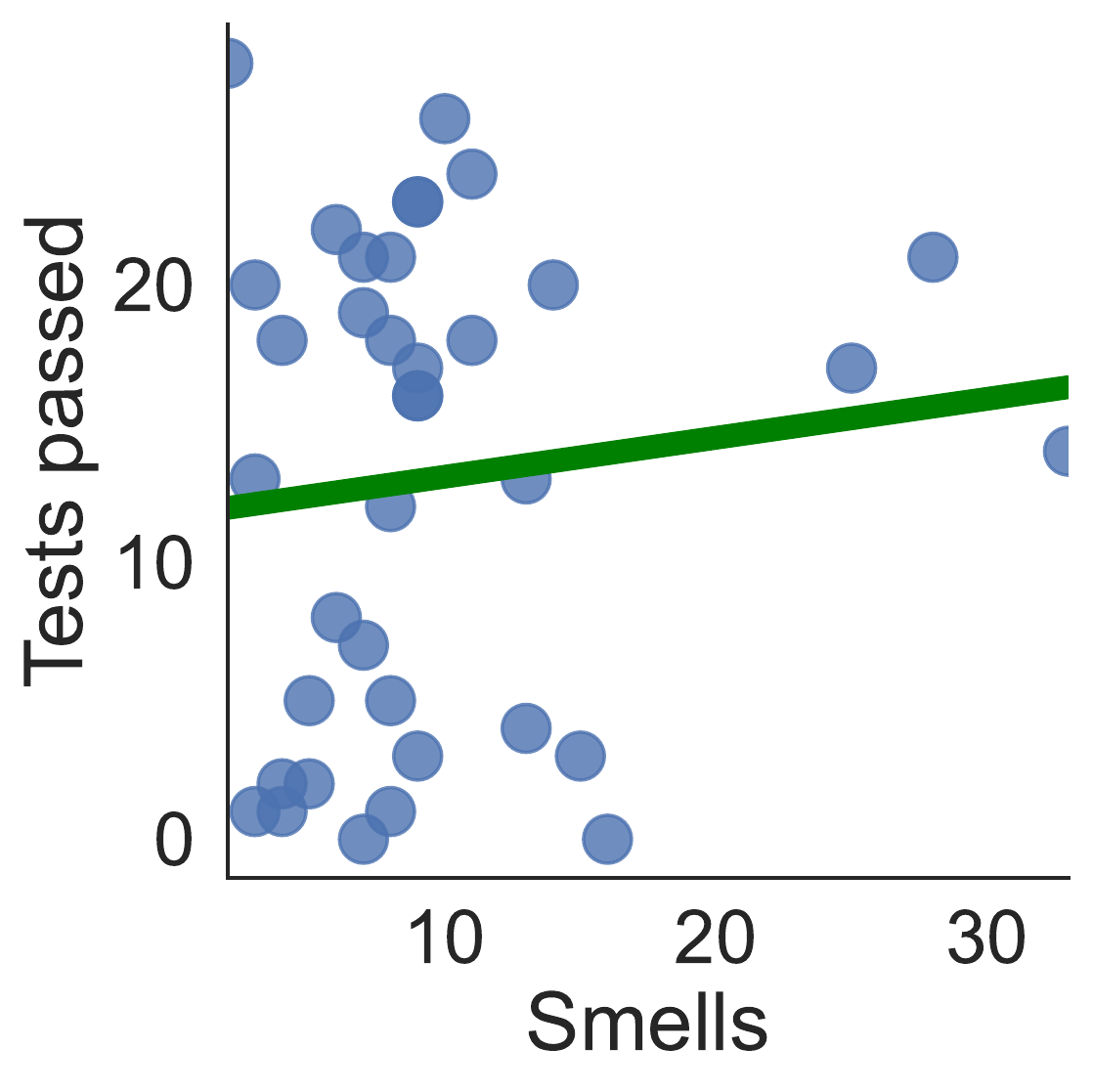}}\hfill
	\caption{\label{fig:blocks_scatter} Relation between tests passed and the number of (a) bug patterns and (b) smells.
	}
\end{figure}

Code smells and bug patterns are alternative means for automated assessment of code.  Whereas the relationship between code perfumes and correctness is clearly visible, the relation between tests passed and the number of bug patterns (Figure~\ref{fig:scatter_bugpattern_correctness}) and number of code smells (Figure~\ref{fig:scatter_smell_correctness}) is not so straight forward.
Our data shows neither a correlation between tests passed with the total occurrence of bug patterns (r=\pearsonPassedPatterns, p=\pvaluePassedPatterns) nor with the number of code smells (r=\pearsonPassedSmells, p=\pvaluePassedSmells). We conjecture that the high $p$-values of the correlation coefficients will be influenced by the substantially lower number of code smells and bug patterns in the student projects compared to code perfumes. Furthermore, an important difference between code perfumes and code smells/bug patterns is that, unlike code perfumes, neither code smells (r=\pearsonSmellBlockCount, p=\pvalueSmellBlockCount) nor bug patterns (r=\pearsonPatternBlockCount, p=\pvaluePatternBlockCount) are correlated to size. 

To compare code perfumes with code smells and bug patterns we therefore control for size through normalisation. This reveals a weak negative correlation between the number of smells per blocks and the number of tests passed (r=\pearsonPassedSmellsBlocks, p=\pvaluePassedSmellsBlocks), and also a weak negative correlation (r=\pearsonPassedPatternsBlocks, p=\pvaluePassedPatternsBlocks) between bug patterns per blocks and the number of tests passed. In contrast, even when normalised the correlation between code perfumes per blocks and the number of tests passed remains higher  (r=\pearsonPassedPerfumeBlocks, p=\pvaluePassedPerfumeBlocks). 
While, considering the strength of the correlations, neither of these analyses can serve as a form of auto-grading of student solutions on its own, it does seem that code perfumes provide a better indication for how good the code is. Furthermore, since code perfumes are orthogonal to code smells and bug patterns, a combination of all three factors is probably best suited to inform teachers about their students' progress.


\summary{RQ2}{Correct solutions tend to contain more code perfumes.}

\section{Conclusions}\label{sec:conclusions}

A common ground is important for providing feedback. When providing feedback about negative aspects of code in \scratch an existing vocabulary of bug patterns and code smells is available. However, in order to provide better, contextualised, and individualised feedback, and especially for encouraging learners, providing feedback about positive aspects of the code is also essential.

In order to provide a vocabulary for good coding practices we introduced and empirically evaluated a catalogue of 25 code perfumes in \scratch, as well as an automated means for detecting them in \scratch programs. Our evaluation found occurrences of all of these code perfume, as well as evidence that code perfumes occur more frequently in correct code.

In this paper we introduced the concept of code perfumes and performed analyses to establish a basic understanding of how common they are and how they relate to correctness. An important next step will be to evaluate the effects this positive feedback has on novice programmers. We furthermore envision that not only learners can benefit from this positive feedback, but also teachers who need to get an overview of their students' \scratch programs in order to provide individualised feedback. Therefore, a further important aspect of future work will be to inspect if the combined use of positive linting (i.e., code perfumes) and negative linting (i.e., code smells and bug patterns) can indeed help teachers provide better feedback.

We introduced an initial set of 25 code perfumes in this paper, but implementing additional code perfumes is straightforward, and we expect to add many more in the future. For example, a desirable perfume could be the usage of meaningful comments inside the \scratch code, but to include such a pattern the characteristics of a good comment in \scratch have to be defined first. 

In order to support the adoption of code perfumes and further research, all our code perfumes are implemented directly into \litterbox, which is freely available at: \url{https://github.com/se2p/LitterBox}.

\begin{acks}
This work is supported by the Federal Ministry of Education and Research
through projects 01JA1924 (SKILL.de) and 01JA2021 (primary::programming) as
part of the ``Qualitätsoffensive Lehrerbildung'', a joint initiative of the
Federal Government and the Länder. The authors are responsible for the content
of this publication.
\end{acks}
\balance

\bibliographystyle{ACM-Reference-Format}
\bibliography{references}

\end{document}